\documentclass[useAMS,usenatbib,referee]{mn2e}
\usepackage{graphicx}

\begin{document}
\title{On the origin of the torus and jet-like structures in the
centre of  Crab Nebula}
\author[Bogovalov \& Khangoulian]
       {S. V. Bogovalov, D.V. Khangoulian \\
       Moscow state Engineering physics institute (technical university)}
\date{ }
\pubyear{2002}

\maketitle

\label{firstpage}

\begin{abstract}
The formation of the toroidal and jet-like structures  in the central
part of the Crab Nebula is explained in the  framework of Kennel \&
Coroniti theory. The only new element introduced by us in this
theory is the initial anisotropy of the energy flux in the wind.
We estimate the   X-ray surface brightness of the Crab Nebula from the
region of interaction of this wind with the interstellar medium
and compare it with observations.
\end{abstract}


\section{Introduction}

The Crab Nebula is powered by the wind of a relativistic $e^{\pm}$
plasma from pulsar PSR 0531+21. The wind is terminated by a shock
front. The particles of the wind are redistributed in energy and
their motion is randomised at the shock. Downstream of the shock
(in the nebula) they emit synchrotron and inverse Compton
radiation \citep{kennel,dejager,aharonian}. Detection of these
emissions is still the only way to obtain information about the
wind (see however \cite{bogah}). Observations in the X-ray
\citep{brinkman,weisskopf} and optical \citep{hester}  have
revealed a remarkable torus as well as jet-like structures in the
central part of the Crab Nebula. The mechanism which produces
these structures apparently gives rise to similar features
observed around the Vela pulsar \citep{pavlov00,pavlov01,helfand},
PSR 1509-58 \citep{kaspi} and in  the supernova remnants G0.9+1
\citep{gaensler} and G54.1+0.3 \citep{lu}. The understanding of
this mechanism will certainly give us new information about pulsar
winds.

The integral characteristics of the Crab Nebula are  described by the
theory of \cite{kennel}. This theory   explains well the spectra
and luminosity of the Crab Nebula in photon energy range from
eV up to TeV gamma-rays \citep{aharonian}. However,
\cite{kennel} strongly simplified the physics of the nebula. They
assumed that  pulsar winds are isotropic. Therefore, this
theory in it's original form is not able in principle to explain
nonuniform structures observed in the Crab Nebula.

Analysis shows that
magnetic collimation of the pulsar winds into jets is impossible in
conventional
theories of the pulsar winds \citep{begelmanli,beskin,bogts}. Therefore,
it is very difficult to interpret the observed jets as the result of
collimation
of the pulsar winds \citep{lubech}.  The observation of the torus
leads to the natural conclusion that
the  acceleration of the wind basically occurs near  the equatorial plane \citep{asbr}.
It was shown recently  \citep{bogkh}(hereafter Paper I) that the formation of the torus and jets
directly follows from
conventional theories of the pulsar winds if the
longitudinal distribution of the energy flux in the wind is taken into
account.
This work is the direct continuation of  Paper I.
Our main goal in this paper   is to estimate  the surface synchrotron
brightness of the central part of the Crab Nebula
and to compare  the results of these estimates  with  observations.

\section{Calculation of the surface brightness}

Pulsar winds have anisotropic distribution of  energy flux
\citep{bog99}. The particle flux can be considered to be more or
less isotropic.
 The key point of our approach is that when this
circumstance is taken into account,
the Lorentz factor of the wind from the Crab pulsar should depend on polar angle $\theta$
as follows  \citep{bogah,bogkh}:
 \begin{equation}
 \gamma=\gamma_0+\gamma_m \sin^2\theta,
 \label{eq1}
 \end{equation}
 here $\gamma_0\approx 200$ is the initial Lorentz factor of the
 wind \citep{daugherty}, and $\gamma_m =
 ({\Omega R_*\over c})^3{B_0^2\over 4\pi n_0 mc^2} \approx 10^6 -10^7$.
 $n_0$, $B_0$ are the initial plasma density  and the
 magnetic field on the surface of the pulsar with radius $R_*$ \citep{bog99}.
The interaction of this wind  with the uniform
interstellar medium results in the formation of a low density ``hot''
region near the equatorial
plane and high density but ``cold'' jet-like features along the rotational axis (Paper I).

In this paper we use the same assumptions as in the paper I. The
most important of them are the following. Calculations of the
synchrotron volume emissivity  in the conventional Kennel \&
Coroniti approach show   that only the X-ray torus should be
detected, while the jet-like features  should be invisible in
X-rays (Paper I). It is easy to understand  why. In the
calculations of the volume synchrotron emissivity $I(r,\theta)$ it
was assumed (as by Kennel \& Coroniti) that the power law spectrum
of electrons is formed at the shock. The evolution of the electron
spectrum is defined by synchrotron cooling. The losses on the
expansion of the nebula are neglected (for details see  Paper I).
All the energy of the electrons injected into the plerion is
radiated as synchrotron radiation. In this case the total
synchrotron luminosity $\int I(r,\theta)r^2dr$ integrated along a
flow line at angle $\theta$ is simply proportional to the energy
density flux injected into the nebula at given angle $\theta$
\citep{ah02}. Thus,  torus-like emission is formed, since the
energy density flux is $\sim \gamma_0+ \gamma_m \sin^2\theta$.
This implies  that the brightness distribution in the torus
depends basically on the energy flux distribution and is less
dependent on the distribution of the particle flux in the pulsar
wind.

We have assumed  an additional acceleration of the electrons in
the volume of the plerion with the amount of the accelerated
electrons proportional to the local density of plasma to brighten
the jet-like features  in the post shock region (Paper I). Some
evidence that the population of the electrons in the ``jets''
really differs from the population of the electrons responsible
for the emission in the torus follows from XMM-Newton observations
\citep{xmm}

The surface brightness is calculated for the inclination angle of the
equatorial plane  in relation to an observer of
$33^{\circ}$ \citep{hester}.

It has been proposed by \cite{pelling} that the high brightness of the
Nort western
part of the torus relative to the low brightness of its opposite part is due to
the relativistic boosting of photons. We took
into account this effect.  The photon flux from different parts of the nebula
has been  transformed  according to the equation \citep{blandford}
\begin{equation}
F(e)=s^{2+\beta}F^{'}(e), \label{boost}
\end{equation}
where $e$ is the photon energy, $s= (\gamma (1-{\bf vn}))^{-1}$,
$\bf n$ is the unit vector directed to the observer, and
$\bf v$ is the local bulk
motion  velocity of the plasma. In the range  $e=400-1000~ \rm
eV$  $\beta$ is close to 2 \citep{aharonian,xmm}.

\section{Comparison with observations}

\begin{figure}
\includegraphics[width=80mm,angle=270.0]{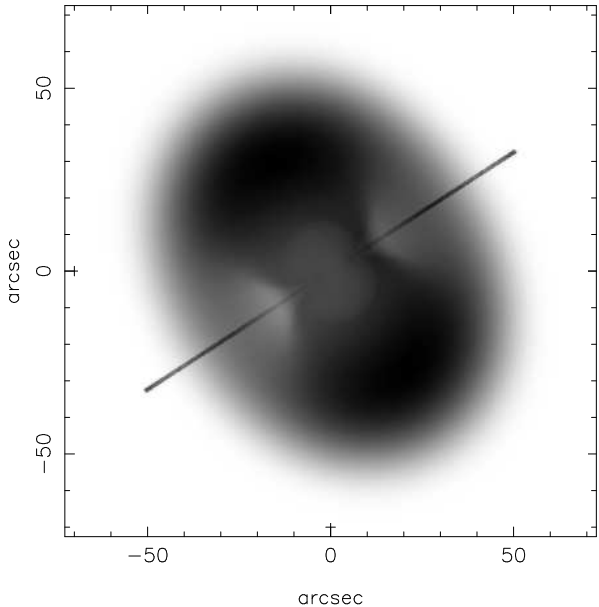}
 \caption{Calculated surface brightness of the nebula.
Relativistic boosting of photons due to the motion of the plasma
in the post shock region results in slight asymmetry of the
brightness in relation to the equatorial plane.} \label{fig1}
\end{figure}

The results of our calculations of the surface brightness of the
nebula are shown in Fig. \ref{fig1}.    It is seen that indeed we obtain the torus of
synchrotron radiation and two jet-like structures. The letter are
visible due to the additional  component of the accelerated particles.

The calculated radial distribution in surface brightness of
the torus  in the equatorial plane corresponds well to that
observed. The maximum in  brightness is reached at a distance of
$\sim 40\arcsec$, which is close to the radius of the outer torus of
$38\arcsec$ \citep{weisskopf}. The calculated torus is evidently
wider than the observed one in the direction perpendicular to the
equatorial plane. This is the first essential difference between
the calculations and observations.

The second difference is that the brightness difference of the
torus edges directed from us and toward us is evidently smaller
than in the observed torus \citep{weisskopf}. The calculated
picture is rather symmetric. Careful inspection of fig. \ref{fig1}
shows that the relativistic boosting only slightly changes the
relative brightness of the opposite edges of the nebula. This
happens because  the regions of maximum brightness  are located at
a distance of $ \sim 3r_{sh}$ where the plasma velocity falls down
below ${c\over 27}$. Relativistic corrections (\ref{boost}) are
rather small here.

We guess that these two disagreements between the calculated
picture and the observed one are of general origin. We assumed
that the flow downstream the shock remains radial.
This is rather crude approximation. Our analysis of the flow
downstream of the shock (Paper I) and direct numerical time-dependent
simulations of the interaction of the anisotropic wind with the
interstellar medium \citep{koldoba} show that the wind is
strongly deflected to the equatorial plane after the shock.
This happens firstly at the
shock. It is easy to make sure that the flow lines diverge to the
equator at the shock front. Downstream of the shock
the flow is further compressed to the equator by the pressure
gradient. It is clear that if this effect is taken into account,
the synchrotron torus will have a smaller width across
the equatorial plane. In addition, the compression of the post
shock flow to the equator will result in a slower decrease of the
velocity with $r$. The effect of the relativistic boosting will be
more stronger. Therefore, we believe that the disagreements
between the calculations and the observations found at the present
stage of our analysis are due to the used approximations. These
disagreements can be reduced or may be  removed altogether through a
 more accurate
treatment of the flow in the post shock region.

The shape of the calculated jet-like
structures differs from the observed ``jets'' as well.
The observed ``jets'' are more laterally extended and
expand with distance from the center. This disagreement is
due to  the assumption
that the additional acceleration of the particles is simply
proportional to the
plasma density in the post shock region. At least two effects will result
in better agreement with the observations. The first one is the deflection
of the flow lines away from the rotational axis as it  follows from real
dynamics of the plasma in the post shock region. The second one  is
an averaging of the density and magnetic field near the
rotational axis at the development of the kink instability,
which must take place
for the configuration of the magnetic field formed in the post shock region
(Paper I).
More accurate treatment of all these processes will result in  better
agreement of the theoretically predicted brightness distribution of the
X-ray Crab Nebula  with observations.

\section{Conclusion}
The fact that the morphology of the  central part of the Crab
Nebula can be explained in frameworks of the theory developed by
\cite{rees,kennel,chevalier} is the basic result of our work. The
elucidation of the nature of the torus and jets
opens for us new horizons.  In particular, comparison of the observed brightness
distribution of the torus with calculations based on an
accurate modelling of the plasma flow in the post shock region will
open the way
to obtain observational information about the energy flux
distribution in the pulsar wind.

\section*{Acknowledgments}
The work is performed under support of collaborative INTAS-ESA
grant N 120-99. We are grateful to Felix Aharonian,  for important
comments. S.V. acknowledges R.Sunyaev,H.Spruit, J.Tr\"umper and
B.Aschenbach for  useful discussion of the problem. The authors
thank Richard Tuffs for help in preparation of the manuscript.


\begin{thebibliography}{99}
\bibitem[\protect\citeauthoryear{Aharonian \& Atoyan}{1998}]{aharonian}
Aharonian F.A., Atoyan A.M. 1998, in Proc. Neutron stars and Pulsars,
Eds. Shibazaki et
 al. Tokyo, Universal acad. Press, Inc., p. 439
\bibitem[\protect\citeauthoryear{Aharonian}{2002}] {ah02}
Aharonian,2002, private communication
\bibitem[\protect\citeauthoryear{Aschenbach \& Brinkmann}{1975}]{asbr}
Aschenbach B., Brinkmann W., 1975, A\&A, 41, 147
\bibitem[\protect\citeauthoryear{Begelman \& Li}{1992}]{begelmanli}
Begelman M.C., Li Z.-Y. 1992, ApJ, 397, 187
\bibitem[\protect\citeauthoryear{Beskin et al.}{1998}]{beskin}
Beskin V.S., Kuznetsova I.V., Rafikov R.R., 1998, MNRAS, 299, 341
\bibitem[\protect\citeauthoryear{Bogovalov \& Aharonian}{2000}]{bogah}
Bogovalov S.V., Aharonian F.A. 2000, MNRAS, 313, 504
\bibitem[\protect\citeauthoryear{Bogovalov}{1999}]{bog99}
Bogovalov S.V., 1999, A\& A, 349, 101
\bibitem[\protect\citeauthoryear{Bogovalov \& Tsinganos}{1999}]{bogts}
Bogovalov S.V., Tsinganos K., 1999, MNRAS, 305, 211
\bibitem[\protect\citeauthoryear{Bogovalov \& Khangoulian}{2002}]{bogkh}
Bogovalov S.V., Khangoulian D.V. 2002, Astronomy Letters, 28, 373
(Paper I)
\bibitem[\protect\citeauthoryear{Brinkman et al.}{1985}]{brinkman}
Brinkmann W., Aschenbach B., Langmeier A.,
1985, Nature, 313, 662
\bibitem[\protect\citeauthoryear{Coroniti}{1990}]{coroniti}
Coroniti F.V., 1990, ApJ, 448, 240
\bibitem[\protect\citeauthoryear{Daugherty \& Harding}{1982}]{daugherty}
Daugherty J.K., Harding A.K., 1982, ApJ, 252, 337
\bibitem[\protect\citeauthoryear{De Jager \& Harding}{1992}]{dejager}
de Jager O.C., Harding A.K., 1992, ApJ, 396, 161
\bibitem[\protect\citeauthoryear{Emmering \& Chevalier}{1987}]{chevalier}
Emmering R.T., Chevalier R.A., 1987, ApJ., 321, 334
\bibitem[\protect\citeauthoryear{Gaensler et al.}{2001}]{gaensler} Gaensler B.M., Pivovarof M.J., Garmire G.P.,
2001, ApJ, 556, 107
\bibitem[\protect\citeauthoryear{Helfand et al.}{2001}]{helfand} Helfand D.J., Gotthelf E.V., Halpern J.P.,
2001, ApJ, 556, 380
\bibitem[\protect\citeauthoryear{Hester et al.}{1995}]{hester} Hester J.J., et al., 1995, ApJ, 448, 240
\bibitem[\protect\citeauthoryear{Kaspi et al.}{2001}]{kaspi} Kaspi V.M., Pivovaroff M.J., Gaensler B.M.,
Kawai N., Arons J., Tamura K., 2001, AAS, 197, 8312
\bibitem[\protect\citeauthoryear{Kennel \& Coroniti}{1984}]{kennel} Kennel C.F., Coroniti F.V., 1984, ApJ, 283, 694
\bibitem[\protect\citeauthoryear{Koldoba \& Ustyugova}{2002}]{koldoba}
Koldoba A., Ustyugova G.V., 2002, privite communication
\bibitem[\protect\citeauthoryear{Lind \& Blandford}{1985}]{blandford}
Lind K.R., Blandfrod R.D. 1985, ApJ, 295, 358
\bibitem[\protect\citeauthoryear{Lyubarsky \& Eichler}{2001}]{lubech} Lyubarsky Yu.E., Eichler D., 2001, ApJ, 562,
p.494
\bibitem[\protect\citeauthoryear{Lu et al.}{2002}]{lu} Lu F.J., Wang Q.D., Aschenbach B., Durouchoux Ph., Song L.M. 2002,
ApJ(L), in press (astro-ph$\setminus$0202169)
\bibitem[\protect\citeauthoryear{Pavlov et al.}{2000}]{pavlov00} Pavlov G.G., Sanwal D., Garmire G.P., Zavlin
V.E., Burwitz V., Dodson R.G., 2000 A\&A, 32, 733
\bibitem[\protect\citeauthoryear{Pavlov et al.}{2001}]{pavlov01} Pavlov G.G., Zavlin V.E., Sanwal D., Burwitz
V., Garmire G.P., 2001, ApJ, 552, L129
\bibitem[\protect\citeauthoryear{Pelling et al.}{1987}]{pelling} Pelling R.M., Paciesas W.S., Peterson L.E. 1987,
ApJ, 319, 416
\bibitem[\protect\citeauthoryear{Rees \& Gunn}{1974}]{rees} Rees M.J., Gunn J.E., 1974, MNRAS, 167, 1
\bibitem[\protect\citeauthoryear{Weisskopf et al.}{2000}]{weisskopf} Weisskopf M.C. et al., 2000, ApJ, 536, L81
\bibitem[\protect\citeauthoryear{Willingale et al.}{2001}]{xmm} Willingale R., Aschenbach B., Griffits R.G., Sembay S.,
Warwick R.S., Becker W., Abbey A.F., Bonnet-Bidaud J.-M., 2001, A\&A, 365, L212
\end{thebibliography}
\end{document}